\documentclass{article}
\usepackage{spconf,amsmath,graphicx, amssymb, hyperref, amsthm, dsfont, amsfonts}
\usepackage{algorithm,algorithmic}
\usepackage{multirow}
\usepackage{multicol}

\long\def\comment#1{}

\newfont{\bbb}{msbm10 scaled 700}

\newfont{\bb}{msbm10 scaled 1000}
\newcommand{\CC}{\mbox{\bb C}}
\newcommand{\RR}{\mbox{\bb R}}

\newcommand{\uv}{{\bf u}}

\newcommand{\vv}{{\bf v}}
\newcommand{\xv}{{\bf x}}

\newcommand{\Am}{{\bf A}}

\newcommand{\Dm}{{\bf D}}

\newcommand{\Id}{{\bf I}}

\newcommand{\Nm}{{\bf N}}

\newcommand{\Tm}{{\bf T}}
\newcommand{\Um}{{\bf U}}
\newcommand{\Wm}{{\bf W}}

\newcommand{\Zm}{{\bf Z}}

\newcommand{\Gc}{{\cal G}}

\newcommand{\herm}{{\sf H}}
\newcommand{\transp}{{\sf T}}

\title{Orthogonal Transforms for Signals on Directed Graphs}

\name{Julia Barrufet$^{1}$, Antonio Ortega$^{2}$}
\address{$^1$Universitat Politècnica de Catalunya, Barcelona, Spain \\ 
	$^2$University of Southern California, Los Angeles, USA}

\begin{document}
\ninept
\maketitle
\begin{abstract}
% The abstract should contain about100 to 150 words
In this paper we consider the problem of defining transforms for signals on directed graphs, with a specific focus on defective graphs where the corresponding graph operator cannot be diagonalized. Our proposed method is based on the Schur decomposition and leads to a series of embedded invariant subspaces for which orthogonal basis are available.  As compared to diffusion wavelets, our method is more flexible in the generation of subspaces, but these subspaces can only be approximately orthogonal. 
\end{abstract}

\section{Introduction}

Graph signal processing (GSP) relies on the availability of a suitable frequency definition and corresponding graph Fourier transform (GFT) \cite{Shuman_2013,Sandryhaila_2013,ortega8347162,Ortega2020GSP}. For undirected graphs, graph operators are symmetric and the GFT is orthogonal. For directed graphs, however, the GFT is not orthogonal. It can be obtained from the eigenvectors of the graph operator, if it is diagonalizable, or using the Jordan form, if not \cite{Sandryhaila_2013}. 
While symmetrization strategies \cite{liu2019solving} can be used to convert directed graphs into undirected graphs or slight modifications can be applied to make a defective directed graph operator diagonalizable (e.g., teleportation in PageRank \cite{pagerankRossi},\cite{Page1999ThePC}), in this paper we focus on the case where it is desirable not to modify the graph operator.

For directed graphs with an adjacency matrix that is not diagonalizable, many studies use the Jordan canonical form  \cite{BRONSON1969176} to find a spectral decomposition \cite{Sandryhaila_2013}\cite{Deri2016GraphSP} \cite{Deri17}. 
%a graph Fourier transform for which the spectral components are the Jordan subspaces of the adjacency matrix is presented. 
More recently, in \cite{misiakos20} a method to replace a given adjacency operator 
%$\Am$
by a diagonalizable operator 
%$\Am_D$ 
obtained via the Jordan-Chevalley decomposition. 
While this method leads to a diagonalizable graph  operator, it starts by computing the Jordan form, which is well known to be numerically unstable. 
Indeed this is a common limitation for all solutions based on the Jordan decomposition. This situation leads to investigate alternative approaches. For example, in \cite{Sardellitti} the graph Fourier basis is built as the set of orthonormal vectors that minimizes a continuous
extension of the graph cut size, known as the Lovasz extension. 
Other alternatives to the Jordan form are presented in \cite{Shafipour_2019} and \cite{Girault_2018}.

%\subsubsection{Modifications for irregular topologies} 

%\subsubsection{Wavelet configurations} 

Finally, an alternative approach would be to use other frequency decompositions as alternatives to the GFT. Examples include the spectral graph wavelet transforms  \cite{HAMMOND2011129}.
%where they propose a novel method for constructing wavelet transforms of functions defined on the vertices of arbitrary finite weighted graphs. Similar ideas are presented in \cite{7088640} and \cite{6557512}.
Our proposed method is inspired by the diffusion wavelets design \cite{coifman2006diffusion} \cite{BREMER200695}. This model offers a tool to build a spectral decomposition for diffusion operators such as the adjacency matrix of a graph. The proposed method, the graph Schur transform (GST), aims to offer a valid operator to perform a spectral decomposition of a graph that can be used even in the case of defective matrices. The numerical advantages of the Schur decomposition were noted in \cite{girault2015signal} and this decomposition was used to obtain block diagonalization of matrices. This approach is used in the GraSP toolbox \cite{girault2017grasp} (see also Appendix B in \cite{Ortega2020GSP}). In this paper we introduce a novel transform design based on the Schur decomposition. 

\section{Preliminaries}
\label{section:schur}
 
%Before presenting the proposed method, we 
%briefly introduce the Schur decomposition, an algebraic method to write a complex matrix as unitarily equivalent to an upper triangular matrix whose diagonal elements are the eigenvalues of the original matrix \cite{PAIGE198111}, and  diffusion wavelets  \cite{coifman2006diffusion}, which shows a possible spectral decomposition of a diffusion operator, such as the adjacency matrix of a graph.

\subsection{Graph signal processing and graph operators}
\label{section:fund}

We define graphs $\Gc(V,E)$ with a set of nodes $V$ and edges $E$, where an edge $e_{ij}$ represents a link between node $i$ and node $j$. A graph can be weighted, if any edge $e_{ij}$ can take a real positive weight $\omega_{ij}$, or unweighted, if the weight for all its edges is 1. We focus on directed graphs where where $e_{ji}$ or $e_{ij}$ may not exist and in general $\omega_{ji}\neq\omega_{ij}$. 
Given a graph $\Gc$ with $N$ nodes, we consider as the main graph operator the adjacency matrix $\Am$, an $N\times N$ square matrix, where the entry $a_{ij}$ will correspond to the weight $w_{ij}$. 
Graph signals are real vectors $\xv\in \RR^N$, where the entry $x(i)$ is the real scalar corresponding to the signal associated to node $i$ and graph filters are constructed as polynomials of the graph operator $\Zm$ (in this paper we focus on $\Zm = \Am$. 
%\subsubsection{Graph filters and operators}
%\label{subsection: filters}
%In GSP, a linear graph filter is represented as a linear operator that gives an output signal $\yv$ when applied to a graph signal $\xv$.
%The most basic example of a linear filter is the adjacency matrix, which gives $\yv=\Am\xv$,
%where the i-th element of the output $\yv$ corresponds to the sum of the values of all its neighbors. 

%This is, the weight of the edge from node $i$ to node $j$. %When the adjacency matrix $\Am$ is defined, we can build the degree matrix of a graph. The \textbf{degree matrix $\Dm$} of a graph is a diagonal matrix where each term of the diagonal corresponds to the total degree of the corresponding node. For an undirected graph $d_{ii}=\sum_j a_{ij}$, adding together all the terms in the corresponding row of the adjacency matrix. For directed graphs we can separate the in-degree matrix and the out-degree matrix, adding together, respectively, the rows and the columns of the adjacency matrix.

For a diagonalizable matrix $\Zm$ we can construct $\Um$, an invertible matrix where each column is one of the eigenvectors of $\Zm$, and write $\Zm=\Um\mathbf{\Lambda}\Um^{-1}$
where $\mathbf{\Lambda}$ is the diagonal matrix with the eigenvalues of $\Zm$, and $\Um$ a matrix containing its eigenvectors. Now we can write any graph signal $\xv$ in terms of its graph frequencies:
\begin{equation}
	\xv=\Um\tilde{\xv} \;\;
\text{where}
\;\;
	\tilde{\xv}=\Um^{-1}\xv
\end{equation}
is defined as the Graph Fourier Transform (GFT) of the graph signal $\xv$ \cite{Sandryhaila_2013}.
In this paper we focus on directed graphs (digraphs) having a defective (non-diagonalizable) adjacency matrix $\Am$. As shown in \cite{barrufet2020graph}, a simple experiment where random Erdõs-Rényi digraphs \cite{Erd_s_2012} are generated demonstrates that sparse random digraphs are very likely to be defective. 
%Since it is possible for an eigenvalue to have algebraic multiplicity greater than 1, we say that $\lambda_i$ is the frequency associated to the subspace $E_i$, which is defined as the span of the linearly independent eigenvectors of eigenvalue $\lambda_i$. 

\subsection{Schur Decomposition}
\label{sec:SD}
\label{subsubsection:sd}
Given $\Zm$ a fundamental graph operator, the Schur decomposition (SD)  \cite{PAIGE198111} of $\Zm$ can be obtained even if $\Zm$ is not diagonalizable as $\Zm = \Um \Tm \Um^{\herm}$ where $\Um$ is unitary, $\Um^{\herm}$ is the Hermitian transpose of $\Um$ and $\Tm$ is an upper triangular matrix with the eigenvalues of $\Zm$ along its diagonal.
The Schur decomposition is not unique, a different one is obtained for any given ordering of eigenvalues.
%eigenvalues with algebraic multiplicity $m_a$ greater than one as is the case for defective $\Zm$ \cite{Ortega2020GSP}.

Denote $\uv_i$ the $i$-th column of $\Um$ and let $E_i = {\rm span}(\uv_i)$. Then, because $\Um$ is unitary, we have that $\CC^N = \bigoplus_{i=1}^{N} E_i$, i.e., $\CC^N$ is the direct sum of the $E_i$ subspaces and it holds that:
\begin{equation}
\label{eq:invariant-space}
	F_k = \bigoplus_{i=1}^{k} E_i, \ k=1 \ldots N
\end{equation}
are subspaces of $\RR^N$ invariant under $\Zm$. 
That is, if $\xv\in F_k$ then $\Zm\xv \in F_k$. Note that since $\Um$ is unitary, we have that $\Um^{\herm} = \Um^{-1}$ so that $\Zm \Um = \Um \Tm$. Thus, defining the upper triangular matrix $\Tm$ as the sum of a diagonal matrix of eigenvalues $\Dm$ and a nilpotent matrix $\Nm$ we have that:
\begin{equation}
\label{eq:invariance-1}
	\Zm [\uv_1  \cdot \cdot \ \uv_N] = [\uv_1 \cdot\cdot \ \uv_N](\Dm + \Nm) =  [\uv_1 \cdot \cdot \ \uv_N]\begin{pmatrix} \lambda_1 & n_{12}  & \ \cdot \ \cdot\\ 0 & \lambda_2 & \ \cdot \ \cdot \\ 0 & 0 & \ \cdot \ \cdot \\ : & : &\cdot \end{pmatrix}
\end{equation}
where $\lambda_1, ...\ , \lambda_N$ represent the eigenvalues of the $\Zm$ operator and $n_{ij}$ the corresponding element of the $\Nm$ matrix. Now the multiplication of each of the $\uv_i$ by the operator $\Zm$ can be expressed as
%\[ \Zm \uv_1 = \lambda_1 \uv_1 \ \in F_1 \]
%\[ \Zm \uv_2 = n_{12} \uv_1 + \lambda_2 \uv_2 \ \in F_2 \]
%and in general
\begin{equation}
	\label{eq:invariance-2}
	\Zm \uv_i = \sum_{j=1}^{i-1} n_{ji} \uv_j + \lambda_i \uv_i \ \in F_i.
\end{equation}
Therefore, subspaces $F_i$ are invariant but there is an overlap, since $F_{i-1}\subset F_i$.  

\subsection{Diffusion Wavelets}
\label{sec:DW}
\label{subsubsection:dw}
Diffusion wavelets (DW)  \cite{coifman2006diffusion} achieve  spectral and vertex domain localization, without compact support in either of these domains \cite{Ortega2020GSP}. The key observation in the DW design is that successive powers of the diffusion operator $\Zm$ will have increasingly lower numerical rank. This holds for a normalized operator $\Zm$, i.e., whose  eigenvalue magnitudes are in the range from 0 to 1, since the eigenvalues of $\Zm^k$ are $\lambda_i^k$ and can become arbitrarily small as $k$ increases. 
%This leads to the definition of $\epsilon$-span of a set of vectors. Consider a set of vectors $ \Phi_v = \{\vv_1, \vv_2, \ldots, \vv_N\}$ which could be for example the columns of $\Zm^k$. Then, define a set of vectors $\Phi_u = \{\uv_1, \uv_2, \ldots, \uv_j\}$ with $j \leq N$. Then $\Phi_u$ $\epsilon$-spans $\Phi_v$ if for all $i=1, \ldots, N$:
%\begin{equation}
%	||\Pm_{\Phi_u}\vv_i -\vv_i||_2 \leq \epsilon  
%\end{equation}
%where $\Pm_{\Phi_u}$  computes the projection of a vector onto the span of $\Phi_u$. Intuitively if  $\Phi_u $ $\epsilon$-span $\Phi_v $  with $j<N$ this means that not much of an error is made by approximating the span with a smaller set of vectors. 
The DW design can be summarized as follows \cite{Ortega2020GSP}. First, select sets: 
\begin{equation}
	\Phi_{\Zm^{2^i}} = \{\lambda_{s,1}^{2^i}\vv_1, \ldots  \lambda_{s,b}^{2^i}\vv_N\}
\end{equation}
where $\{\vv_1, \ldots  \vv_N\}$ are the eigenvectors of $\Zm$. Each of these sets corresponds to the eigenvalues of the matrices obtained from consecutive dyadic powers of the operator $\Zm$ such that $\lambda<\epsilon$. Then $V_0 = \RR^N$ and $V_i = {\rm span} (\Phi_i)$, where $\Phi_i$ $\epsilon$-spans  $\Phi_{\Zm^{2^i}}$ \cite{coifman2006diffusion}. At any stage we then find a subspace $W_i$ such that 
\begin{equation}
	V_i \oplus W_i = V_{i-1}. 
\end{equation}
Then choosing a specific $i$ we write
\begin{equation}
	V_0 = V_i \oplus W_i \oplus W_{i-1} \oplus \ldots \oplus W_1
\end{equation}
so that the basis for the $W_i$ spaces form the orthogonal wavelets and the basis for $V_i$ correspond to the scaling function. From this design we can finally obtain $M+1$ orthogonal subspaces formed by orthogonal vectors and corresponding to the eigenvalues of $\Zm$ in a increasing order $\{W_1,\ W_2, \ ...\ , W_M, \ V_M\}$.
DW has many advantages for directed graphs, including cases where $\Zm$ is defective, since the basis obtained are orthogonal and an eigendecomposition of $\Zm$ is not required  \cite{coifman2006diffusion} and \cite{BREMER200695}. However, this design is not flexible due to its dyadic structure (powers of two in $\Zm^{2^i}$), which may lead to difficulties when the eigenvalue distribution is highly irregular. 
%. This is why in the next section we propose a new method, with the intention to offer a slightly different outcome, specifically in relation to the invariance of its subspaces and the regularity on the defined distribution.

\section{Graph Schur Transform (GST)}

\label{sec:GST}

%\subsection{Subspace decomposition} 

Both, DW and the SD decompose the space $V_0 = \CC^N$ into the direct sum of orthogonal subspaces. For SD $\CC^N = \bigoplus_{i=1}^{N} E_i$,
%=E_1 \oplus E_2 \oplus ... \oplus E_N$, 
where the subspaces $E_i = {\rm span}(\uv_i)$ are orthogonal, but they are not invariant. Instead, we have that the $F_k$ spaces in \eqref{eq:invariant-space} 
%$= \bigoplus_{i=1}^{k} E_i$ 
are invariant, but clearly they are not orthogonal to each other since $F_{k-1} \subset F_k$.

In the DW design, for a sufficiently large $i$, $V_i$ contains only signals that are in the subspace corresponding to the largest eigenvalue, while $W_1$ corresponds to the smaller eigenvalues, and for increasing levels the subspace $W_i$ corresponds to higher eigenvalues. Also, in DW  subspaces are orthogonal by construction but they are not exactly invariant to multiplication by the graph operator.
Note that if the eigenvalues are ordered in decreasing order and $i$ is sufficiently large, then $F_1 = E_1 = V_i$. 

%Therefore, a strong relation between both methods may be possible, and a decomposition of $\Am$ into subspaces can be constructed based on the Schur Decomposition. 
%This new approach can offer advantages compared with the DW method. First, instead of an approximate invariance  (as the one that gives the Diffusion Wavelets approximation for vectors in the $W_1$ subspaces), the proposed method ensures that vectors within the computed basis are exactly invariant to the subspace they belong to. Also, invariant subspaces can be built in a more regular and flexible way depending on the purpose of the study, in contrast to the arbitrary grouping of eigenvalues that is obtained with the Diffusion Wavelets method.

\subsection{GST Construction} 

The proposed GST is built on the construction of a set of invariant subspaces derived from the SD.
Start with a normalized operator $\Zm$, e.g., the adjacency matrix $\Am$ divided by the magnitude of its largest eigenvalue \cite{Sandryhaila_2013}. 
Given the embedded spaces $F_1 \subset F_2 \subset \ldots \subset F_k$ from SD, we introduce a series of subspaces $G_1,...,G_k$, such that $G_1 = F_1$,   $G_1 \oplus G_2 = F_2$ and, in general,
\begin{equation}
	G_{i-1} \oplus G_i = F_i.
\end{equation}
$G_1$ contains the basis in SD associated to eigenvalues $\lambda_1$ to $\lambda_{i_1}$, but $G_2$ contains a basis corresponding to the eigenvalues from $\lambda_{i_1+1}$ to $\lambda_{i_1+i_2}$ (which  will be renamed $\lambda_1^{(2)}$ to $\lambda_{i_2}^{(2)}$). Then to form a complete basis for $\RR^N$ we can represent this space as:
\begin{equation}
	G_1 \oplus G_2 \ldots \oplus G_k = F_k = \RR^N
\end{equation}
where $F_k$ corresponds to the last subspace of dimension $N$.

%To find the most appropiate criteria to apply to separate the graph frequencies into subspaces, many options have been considered, looking for the one whose output offers better properties on subspace invariance and orthogonality. The reason why the procedure shown below has been chosen will be developed in further detail in the next chapter.

%\subsubsection{Building the subspaces}

From the SD $\Zm = \Um \Tm \Um^\herm$  we can define $G_1$ as:
\begin{equation}
	G_1 = {\rm span}(u_1, ... , u_{i_1}),
\end{equation}
where $\uv_1,\uv_2,...$ correspond to the first columns of the matrix $\Um$ and $\uv_{i_1}$ the column corresponding to the last eigenvalue included in the first subspace. Therefore, the basis for this subspace will be 
\begin{equation}
	\Um_1 = [\uv_1 \  \uv_2 \ ...\  \uv_{i_1}]
\end{equation}    
Due to the orthogonal nature of $\Um$ we have obtained an orthogonal invariant basis for the vectors in $G_1$.

\begin{figure}[h]
	\centering
	\includegraphics[width=\linewidth]{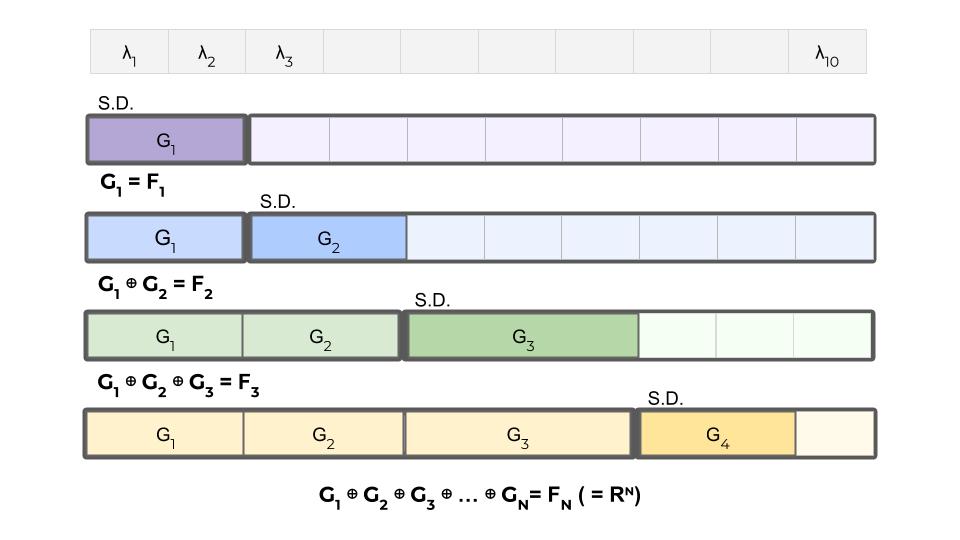}
	\caption{Diagram showing how subspaces are created in the GST method for an example with N=10 where $\lambda_1,\lambda_2,$ etc represent the normalized eigenvalues.}
	\label{fig:GST}
\end{figure}

To build the next subspace, we start by reordering the eigenvalues as $\{\lambda_{{i_1}+1}, \ \lambda_{{i_1}+2}, \ ... \ , \lambda_1, \lambda_2, \ ... \ , \lambda_{i_1}\}$ and renaming the eigenvalues as $\{ \lambda_1^{(2)},..., \lambda_N^{(2)}\}$.
Now we can rebuild the Schur matrix in the new order:
\begin{equation}
	\Tm^{(2)} = 
	\begin{pmatrix} 
		\lambda_1^{(2)} & n_{12}^{(2)} & \cdot & n_{1N}^{(2)} \\ 
		0 & \lambda_2^{(2)} & \cdot & n_{2N}^{(2)}  \\
		\cdot &\cdot &\cdot &\cdot &\\
		0 & 0 & \cdot & \lambda_N^{(2)} 
	\end{pmatrix}
	= \Um^{(2)\transp} \Zm \Um^{(2)}
	\label{eq:SD-iteration}
\end{equation}
and repeating the procedure used for $G_1$, we can build a basis for the space $G_2$ taking the first $i_2$ columns of the matrix $\Um^{(2)}$. The first column will be an eigenvector for $\lambda_1^{(2)}$ and the other $i_2-1$ vectors will be invariant to the subspace they form. With these vectors we will have an orthogonal invariant basis, $G_2$, which will form the orthogonal matrix $\Um_2$:
\[
\Um_2=[\uv_1^{(2)} \ \uv_2^{(2)} \ . . . \ \uv_{i_2}^{(2)} ]
\]
At this point we have built 2 different orthogonal and invariant basis for two sets of energies of the graph: $\lambda_1 ,..., \lambda_{i}$ and $\lambda_1^{(2)},...,\lambda_{i_2}^{(2)}$. Repeating this procedure as many times as necessary will result in a set of $k$ invariant subspaces $G_1,...,G_k$ corresponding to the $k$ groups of energies of the graph, where 
\begin{equation}
	G_1\oplus ... \oplus G_i=F_i \text{, } 
\end{equation}
\begin{equation}
	F_{i}\oplus G_{i+1} = F_{i+1} \text{ and } F_k = \RR^N
\end{equation}
Finally, we can define the graph Schur transform (GST) $\Um_S$ (Fig.~\ref{fig:GST}):
\begin{equation}
\label{eq:GST}
	\Um_S = [\Um_1 \ \Um_2 \ \Um_3 \ ... \ \Um_k].
\end{equation}
Eigenvalues can be ordered by increasing magnitude ($|\lambda|$). 
Given a desired number of subspaces $k$  the eigenvalues $\lambda_{i_1},\lambda_{i_2},...$, which are the end points of each of the subspaces, can be found according to different criteria, e.g., ensuring the distance between consecutive eigenvalues $|\lambda_{i_a},\lambda_{i_{a}+1}|$ is large enough (as discussed next). 

%This configuration computes a set of subspaces $G_i$ invariant to $\Zm$ build with unitary and orthogonal vectors. Therefore, all the basis for $G_1,...,G_k$ are orthonormal and invariant. However, vectors from different subspaces may not be orthogonal to each other, so the final basis $\Um_S$, which will represent the GST of the graph, is not an orthogonal basis.
%This property will be further explained in Section \ref{subsection:orth}.

\subsection{GST properties}
\label{section:prop}

%Basis built both with the Graph Schur Transform (GST) method and the Diffusion Wavelets (DW) have invariance and orthogonality properties. In this section we are going to go over the most interesting properties of the developed method (GST), and compare it with the DW idea.

%\label{subsection:invariance}
%While the basis built in DW present an approximated invariance, the proposed method shows that subspaces are exactly invariant.

%\subsubsection{Invariant subspaces for dyadic powers of $\Am$ in DW}
%In the case of DW, defining $\xv$ as a vector from the subspace $W_1$ built in the first iteration and corresponding to eigenvalues of $\Am$ such that $\lambda<\epsilon$, it holds that $\tilde{\Am}\xv\approx 0$. For a vector $\xv \in W_2$, we have that $\tilde{\Am}^2\xv\approx 0$, for $\xv \in W_3$, $\tilde{\Am}^4\xv\approx 0$, etc. An in general, for $\xv\in W_i$, $\tilde{\Am}^{2^{i-1}}\xv\approx 0$. So subspaces in this method are approximately invariant.

%\subsubsection*{Invariance}

\par 
\par\noindent 
\textbf{Invariance} \quad 
Any  $\Um_i$ in $\Um_S$ from \eqref{eq:GST} obtained with a SD in the form of \eqref{eq:SD-iteration}. As shown in Section \ref{sec:SD}, from \eqref{eq:invariance-1} and \eqref{eq:invariance-2} the corresponding space $G_i$ is invariant to multiplication by $\Zm$. That is, for any $\xv \in G_i$, $\Zm \xv \in G_i$. 

\par
\par\noindent 
\textbf{Orthogonality} \quad 
For both DW and SD subspaces are constructed with  with orthonormal bases of vectors, so that for the GST we have that $\Um_i^\transp\Um_i=\Id$, and for DW it holds that $\Wm_i^\transp\Wm_i=\Id$. In DW, subspaces are orthogonal to each other by construction. Therefore,  $W_i^\transp W_j=\boldsymbol{0}$, and also that $V_k^\transp W_i=\boldsymbol{0}$.
%, since
%\[
%V_0 = V_k \oplus W_k \oplus W_{i-1} \oplus W_{i-2} \oplus \ldots \oplus W_1
%\]
%where all the subspaces $W_i$ are orthogonal to each other and also orthogonal to the subspace $V_k$. 
%Therefore, it holds that $W_i^\transp W_j=\boldsymbol{0}$, and also that $V_k^\transp W_i=\boldsymbol{0}$.
%\subsubsection{Approximate orthogonality for basis in the GST}
In the case of the GST, we have that $\Um_i^\transp \Um_j=\boldsymbol{0}$ does not necessarily hold. However, we can obtain subspaces that are close to  orthogonal by careful selection of the number and boundary between subspaces. Empirically, we observe that choosing a smaller number of subspaces with larger spectral gaps ($|\lambda_{i_a},\lambda_{i_{a}+1}|$) between them tends to favor orthgonality \cite{barrufet2020graph}.

\par
\par\noindent 
\textbf{Spectral Localization} \quad 
The proposed method offers, compared to DW, a more regular and localized distribution of frequencies into subspaces. When building the GST we can decide the exact number of basis created, so that the dimensions of subspaces can be adjusted according to the purpose of the processing needed and also separated in ranges with similar magnitude. This provides added flexibility and enables multiple GST transform designs to be used for a given graph.   %characteristic not only gives more flexibility to our method compared to others, it also allows us to easily create countless variations of the method, adaptable to the interest of the study.

\section{Example GST Construction and experiments}

%This computation would correspond to the 4th step of Algorithm \ref{alg:sw}, and is described in more detail in the following algorithm:
%\begin{algorithm}[H]
%	\caption{Distribution of eigenvalues into subspaces from the GST} 
%	\begin{algorithmic}[1]
%		\STATE $\ev \leftarrow$ Sort the set of eigenvalues of $\tilde{\Am}$
%		\STATE $\dv \leftarrow$ Calculate and store the distance between consecutive eigenvalues.
%		\STATE $\bv \leftarrow$ Find the $M-1$ higher values of $\dv$.
%		\STATE $\kv \leftarrow$ Store the position of the $M-1$ higher values, to know the separation points between subspaces.
%	\end{algorithmic}
%	\label{alg:eig}
%\end{algorithm}
%\vspace{10pt}

\par
\begin{figure*}
		\centering
		\vbox{
		\hbox{\includegraphics[width=0.51\linewidth]{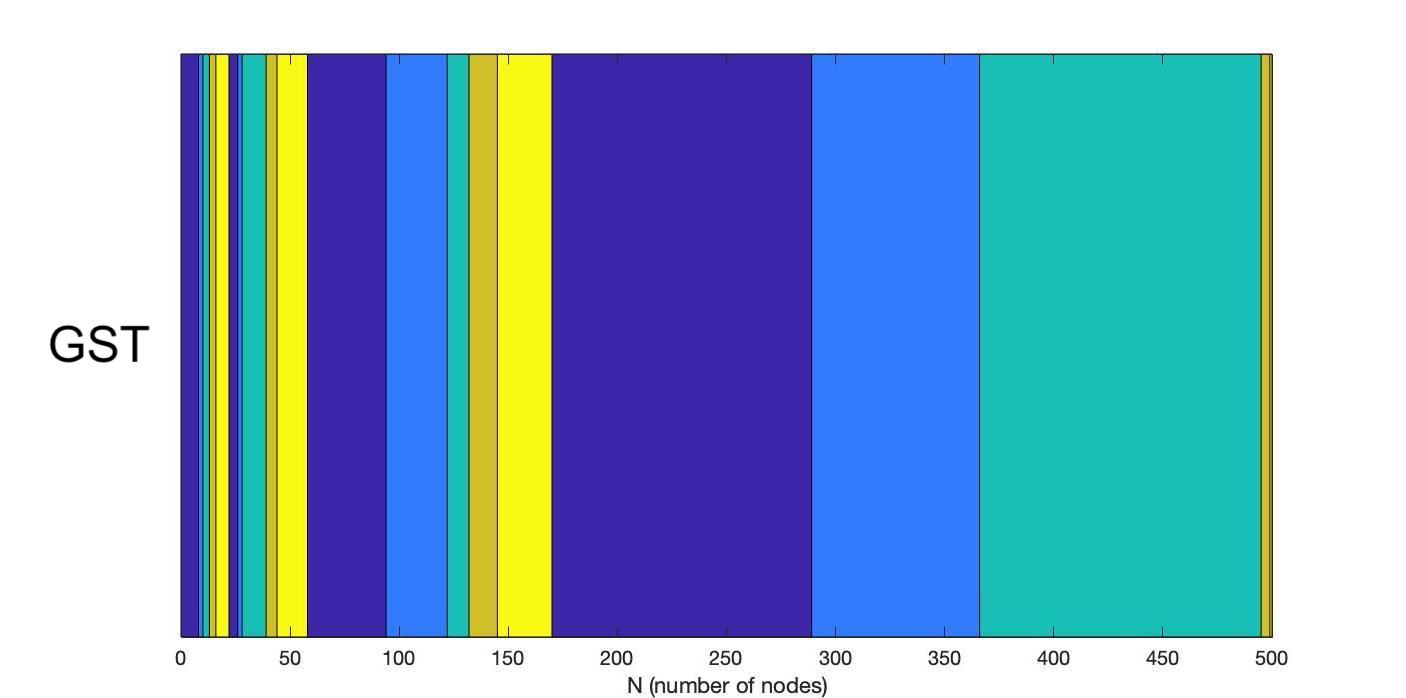}
		\includegraphics[width=0.51 \linewidth]{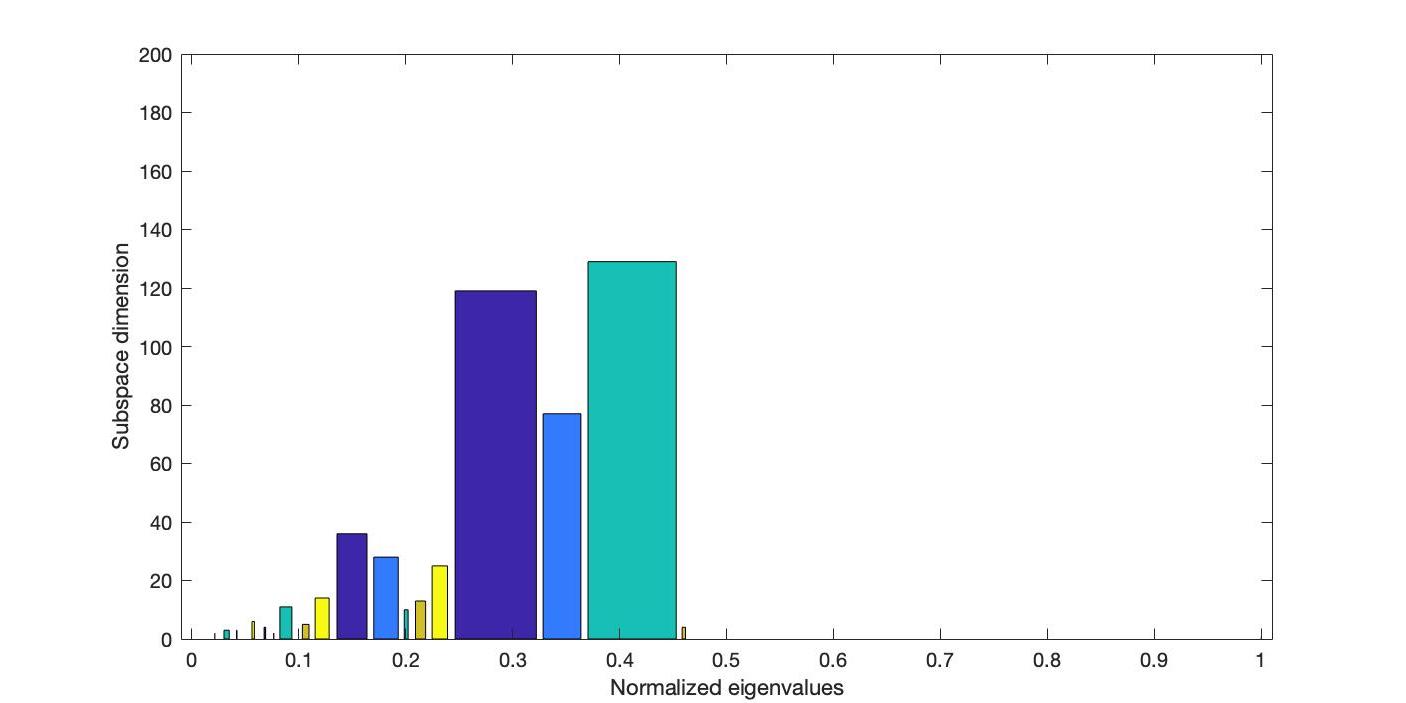}}
		\hbox{\includegraphics[width=0.51 \linewidth]{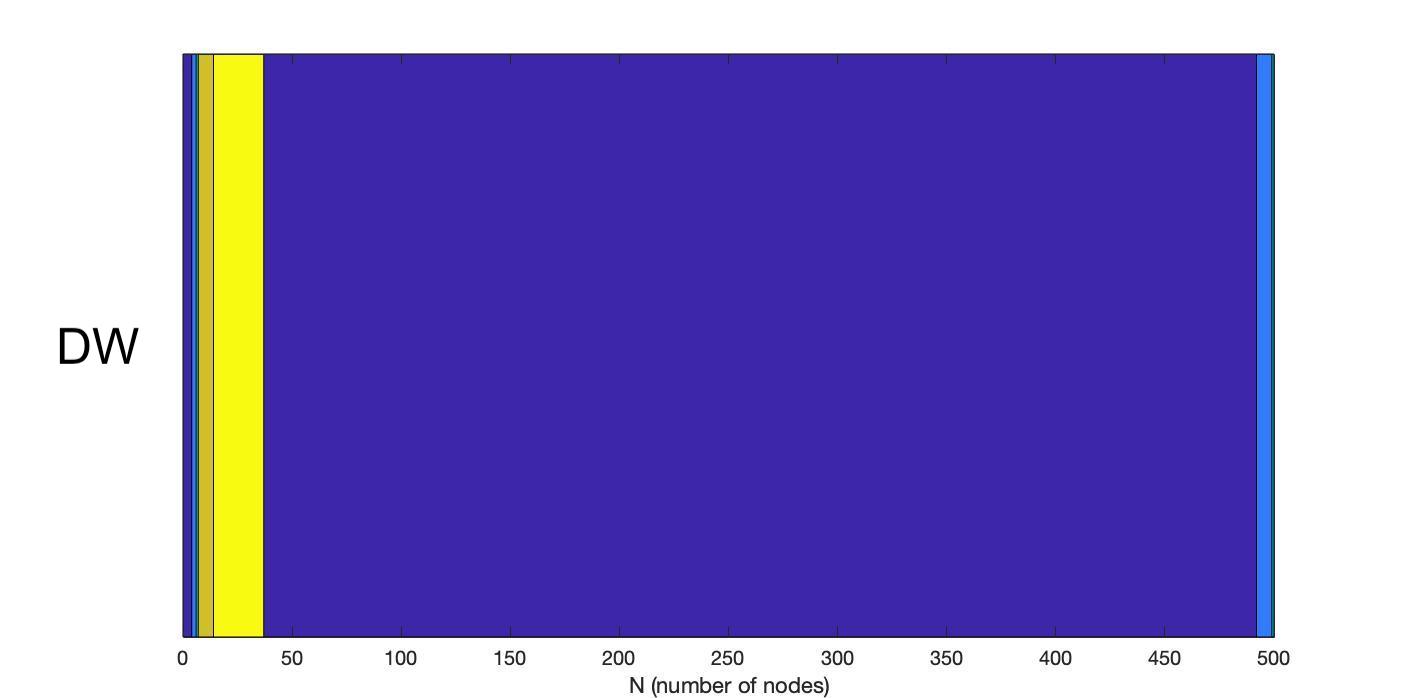}
		\includegraphics[width=0.51 \linewidth]{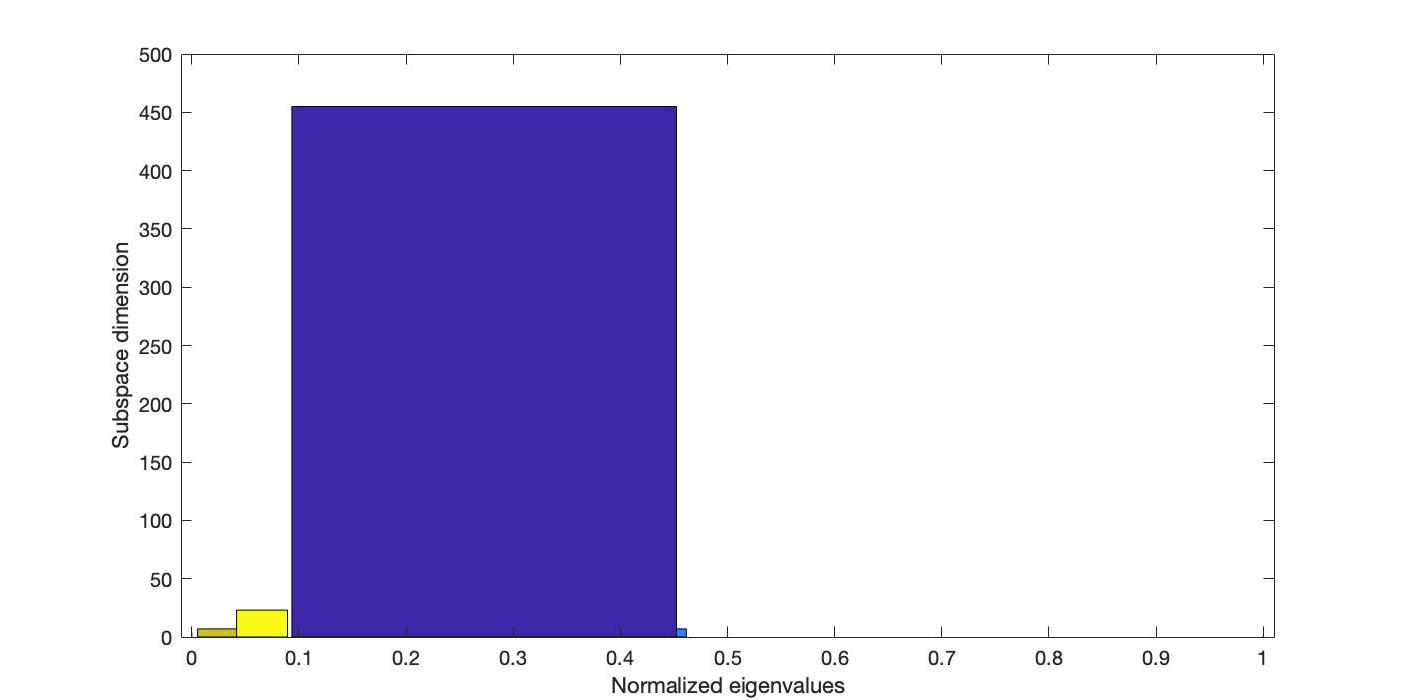}}
		}
	\caption{Left: For GST (top) and DW (bottom) each plot shows how the 500 eigenvalues of the graph are grouped into subspaces. Notice that GST generates 30 subspaces while DW is limited to fewer, with one very high dimensional space. Right: corresponding plots with the range of normalized eigenvalues in each subspace and the subspace dimension.}
	\label{fig:compare}
\end{figure*}

\noindent
\textbf{Dimensions of subspaces in DW and GST} \quad 
To illustrate the proposed method, assume the only parameter to decide is the desired number of subspaces $M$. Then, as a criterion to group eigenvalues into different subspaces will be to find the $M-1$ separation points where the distance between consecutive eigenvalues is greater. 
We test GST and DW for a random synthetic graph of 500 nodes setting the tuning parameter $M=30$ for GST and the $\epsilon$-span parameter $\epsilon=10^{-5}$ for DW (see Section~\ref{sec:DW}).  
For this graph, the distribution of eigenvalues into subspaces is shown in Fig. \ref{fig:compare}. It is clear from the these results that the GST shows a better spectral localization with a smaller variance in the magnitude of eigenvalues contained in each subspace, providing a more precise handling of the subspaces.

\par\noindent 
\textbf{GST for directed graphs} \quad 
The ultimate goal of this research was to find a method to build a basis valid to use GSP tools on directed graphs, and especially on those with a defective adjacency matrix.
%As explained in Section \ref{section:dig} the reason why exceptions cannot happen on undirected graphs is because they are, by definition, always diagonalizable. That is, that a complete basis of eigenvectors can be build by diagonalizing any of its graph operators (in our case, the adjacency matrix $\Am$). This basis forms a transformation matrix $\Um$ such that
%\[ \Am=\Um\mathbf{\Lambda}\Um^{-1} \]
%where $\mathbf{\Lambda}$ is a diagonal matrix formed by the eigenvalues of $\Am$. The presented $\Um$ matrix is needed to build the GFT of any graph signal, as introduced in chapter \ref{chapter:fund}, where
%\[ \tilde{\xv}=\Um^{-1}\xv \]
%
%In the case of directed graphs the adjacency matrix is not always diagonalizable and other graph operators such as the Laplacian cannot be computed because of the nature of the graph and the presence of sinks and sources. Our main purpose was to find a solution for this kind of graphs.
%
For a non-diagonalizable or defective graph a complete set of linearly independent eigenvectors does not exist, so a complete basis of eigenvectors for $\RR^N$ cannot be formed. 

%This means that, even the $\Um$ matrix can be computed, its rank will be ${\rm ran}(\Um)<N$ and this would make the matrix $\Um$ not singular. Therefore the matrix $\Um$ cannot be used to form the GFT of a signal in graphs with a defective $\Am$ matrix, and our proposed matrix $\Um_f$ can be used instead.

%\subsection{Numerical results on directed graphs}

To evaluate experimentally test the proposed method, we 
generate $\Um_S$ matrices with the GST method for random synthetic graphs. We  generate graphs of different sizes and select a different number of subspaces in each case.  In particular, the performed verification consisted on the following tests:
\begin{itemize}
	\item Graphs of 50, 100, 150, 200, 300 and 500 graphs were created.
	\item For each size, the GST was computed for different numbers of subspaces ($M$): $N/25$, $N/10$, $N/5$ and $N/2$.
	\item All graphs where created with a random edge probability $p\in [5/N, 1/N]$.
	\item For each selected $N$ and $M$, 100 graphs were created to calculate the proportion of non-diagonalizable graphs.
\end{itemize}

\begin{figure}[H]
	\centering
	\includegraphics[width=\linewidth]{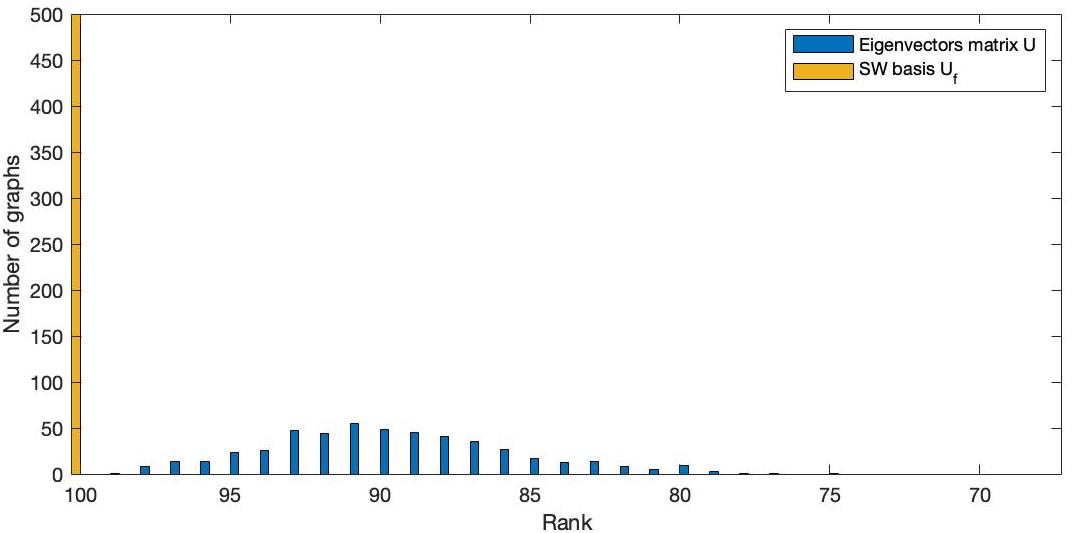}
	\caption{Histogram showing the rank of the Schur $\Um_S$ matrix (yellow) and the matrix of eigenvectors $\Um$ (blue) for 500 graphs with N=100. The graphic clearly shows that, while the adjacency matrix was defective for the 500 tested graphs, the GST always forms a complete basis for $\RR^N$, with ran($\Um_f$)$=N$.}
	\label{fig:rank}
\end{figure}

For each graph we construct a matrix $\Um_e$ with all the eigenvectors for the adjacency matrix of the graph. The adjacency matrix was defective in all cases, with a number of eigenvectors averaging around 90 (Fig. \ref{fig:rank}), while as expected the matrix $\Um_S$ was fullrank for all graphs and indeed provides a set of basis vectors for the space of graph signals.  

%This shows that for any directed graph, either strongly or weakly connected, our method builds a complete basis of $\RR^N$ that can be used to study the graph in the frequency domain. 

Note that in the case of directed acyclic graphs our method will not perform correctly for various reasons. First, the adjacency matrix $\Am$ cannot be divided by the maximum eigenvalue since, by definition, all the eigenvalues are zero for a DAG. Second, the adjacency matrix for a DAG is a nilpotent matrix, that is, an upper triangular matrix with zeros (the eigenvalues) in its diagonal. Therefore, we have that the Schur matrix $\Tm$ has the same form as the adjacency matrix $\Am$. This means that in this case the transformation matrix $\Um$ in the expresion $\Am=\Um\Tm\Um^\herm$ would be $\Um=\Id$. Which makes it not useful for our purpose. 

Finally, the construction of our method groups eigenvalues into $M$ groups, separating them when their magnitudes are different enough. This is not possible in the case of DAGs, where all the eigenvalues have magnitude zero.

%% Añadir grafos MUY excepcionales por ejemplo todos los autovalores 0 y 1 (sinks,sources y ciclo) como la matriz A_

\section{Conclusions}

Our proposed method has several useful properties, including invariance, orthogonality within each subset and flexibility in the subset construction, including the option of constructing near orthogonal subspaces. Our future work may focus on a more thorough evaluation of the potential benefits of GSTs in applications as well further study of techniques that may be valid for directed acyclic graphs, where GST cannot be used.

\newpage
\bibliographystyle{IEEEbib}
\bibliography{refs}

\end{document}